\documentclass[sigconf,screen]{acmart}

\AtBeginDocument{%
  }

\copyrightyear{2024}
\acmYear{2024}
\setcopyright{acmlicensed}
\acmConference[CHI PLAY Companion '24]{Companion Proceedings of the Annual Symposium on Computer-Human Interaction in Play}{October 14--17, 2024}{Tampere, Finland}
\acmBooktitle{Companion Proceedings of the Annual Symposium on Computer-Human Interaction in Play (CHI PLAY Companion '24), October 14--17, 2024, Tampere, Finland}
\acmDOI{10.1145/3665463.3678866}
\acmISBN{979-8-4007-0692-9/24/10}

\usepackage{float}
\usepackage{balance}

\usepackage[shortlabels]{enumitem}

\usepackage{xspace}

\usepackage{microtype}  %define letter spacing
\newcommand{\codestyle}[1]{\textit{#1}}
\newcommand{\mechanics}[1]{\texttt{\textls[-40]{#1}}}

\usepackage{multirow}
\usepackage{fontawesome}

\usepackage{colortbl}
\makeatletter
\def\adl@drawiv#1#2#3{%
	\hskip.5\tabcolsep
	\xleaders#3{#2.5\@tempdimb #1{1}#2.5\@tempdimb}%
	#2\z@ plus1fil minus1fil\relax
	\hskip.5\tabcolsep}
\newcommand{\cdashlinelr}[1]{%
	\noalign{\vskip\aboverulesep
		\global\let\@dashdrawstore\adl@draw
		\global\let\adl@draw\adl@drawiv}
	\cdashline{#1}
	\noalign{\global\let\adl@draw\@dashdrawstore
		\vskip\belowrulesep}}
\makeatother

\usepackage{soul}

\usepackage{xcolor,colortbl}
\definecolor{palered}{HTML}{ffadad}
\definecolor{paleorange}{HTML}{ffd6a5}
\definecolor{paleyellow}{HTML}{fdffb6}
\definecolor{palegreen}{HTML}{caffbf}
\definecolor{palecyan}{HTML}{9bf6ff}
\definecolor{paleblue}{HTML}{a0c4ff}
\definecolor{palepurple}{HTML}{bdb2ff}
\definecolor{palepink}{HTML}{ffc6ff}
\definecolor{palewhite}{HTML}{fffffc}

\definecolor{redwood}{HTML}{B35662}

\definecolor{ferngreen}{HTML}{00A300}

\definecolor{skyblue}{HTML}{219EBC}

\definecolor{amethyst}{HTML}{8758C6}

\definecolor{UTorange}{HTML}{FB8500}

\usepackage{rotating}

\begin{document}

\title[Culture Clash]{Culture Clash: When Deceptive Design Meets Diverse Player Expectations} %The Role of Cultural Attributes in Deceptive Game Design. we need a cool title, something like "Slay the Dragon"

\author{Hilda Hadan}
\authornote{These authors contributed equally to this work.}
\email{hhadan@uwaterloo.ca}
\orcid{https://orcid.org/0000-0002-5911-1405}
\affiliation{
    \institution{Stratford School of Interaction Design and Business, University of Waterloo}
    \city{Waterloo}
    \country{Canada}
}

\author{Sabrina A. Sgandurra}
\authornotemark[1]
\email{sasgandu@uwaterloo.ca}
\orcid{https://orcid.org/0000-0003-3187-263X}
\affiliation{
    \institution{English Language and Literature, Faculty of Arts, University of Waterloo}
    \city{Waterloo}
    \country{Canada}
}

\author{Leah Zhang-Kennedy}
\email{lzhangke@uwaterloo.ca}
\orcid{https://orcid.org/0000-0002-0756-0022}
\affiliation{
    \institution{Stratford School of Interaction Design and Business, University of Waterloo}
    \city{Waterloo}
    \country{Canada}
}

\author{Lennart E. Nacke}
\email{lennart.nacke@acm.org}
\orcid{https://orcid.org/0000-0003-4290-8829}
\affiliation{
    \institution{Stratford School of Interaction Design and Business, University of Waterloo}
    \city{Waterloo}
    \country{Canada}
}

\renewcommand{\shortauthors}{Hilda Hadan, Sabrina Alicia Sgandurra, Leah Zhang-Kennedy, \& Lennart E. Nacke}

\begin{abstract}
Deceptive game designs that manipulate players are increasingly common in the gaming industry, but the impact on players is not well studied. While studies have revealed player frustration, there is a gap in understanding how cultural attributes affect the impact of deceptive design in games. This paper proposes a new research direction on the connection between the representation of culture in games and player response to deceptive designs. We believe that understanding the interplay between cultural attributes and deceptive design can inform the creation of games that are ethical and entertaining for players around the globe.

%\hilda{current word count = 95. Limit: 100-word max}
\end{abstract}





\maketitle

%==============
\section{Introduction}
\label{sec:Introduction}
%\hilda{perspectives of play, deadline = June 5 page limit : 3 pages max, non-anonymous}
%submission cite:~\url{https://chiplay.acm.org/2024/perspectives-on-play/}

%\hilda{A brief rationale (2-3 sentences in response to a PCS system question) on how the contribution is not only of interest to the CHI PLAY community, but also provocative. A 250-word biography of the author(s)/presenter(s) (in the PCS system)}

%==========================
%\hilda{talk about deceptive design in general in 1-2 sentences. Then introduce deceptive game design}

Deceptive design, the design practices that subvert, impair, or distort users' ability to make autonomous and informed decisions~\cite{gray2024ontology}, is prevalent across websites~\cite{mathur2021makes,gray2023mapping,Brignull2023book}, mobile apps~\cite{lewis2014irresistible,fitton2019F2P,gunawan2021comparative}, video games~\cite{roffarello2023defining,zagal2013dark,king20233d}, robotics~\cite{lacey2019cuteness} and immersive environments~\cite{king20233d,hadan2024deceived,krauss2024what}. In games, deceptive design leads to negative player experiences, such as impulsive purchases, longer gameplay sessions, and psychological, social, and emotional dissonance~\cite{zagal2013dark,hhadan2024ow2,gray2021enduser,maier2019dark}, while benefiting the developer and publisher~\cite{zagal2013dark}. Although existing literature explores the characteristics, functions, and impacts of deceptive designs~\cite{gray2024ontology,mathur2021makes,gray2021enduser} and the ethical consideration of manipulating user behaviours~\cite{geronimo2020UI,gray2018dark,lacey2019cuteness}, the focus has primarily been on game mechanics and monetization strategies. Currently, no studies within this area have examined the role of cultural heritage and references in deceptive game design practices. 

%==========================
%\hilda{talk about cultural and attributes in games, why it is important in game desing, and bring connection to player engagement and perception, then connect to potentional malicious use---player manipulation}

\textit{Culture} represents the unique ``spiritual, material, intellectual, and emotional traits of a society,'' including art, architecture, language, values, traditions, and beliefs~\cite[p.~9]{riviere2009investing,tylor1871primitive}. Implementing cultural representations is crucial for designing products that are successfully usable by global users~\cite{heimgartner2013intercultural}. Research shows that players find games incorporating their own culture more enjoyable and attractive~\cite{Pyae2018engagement}. As culture shapes identity, incorporating cultural representations allow players to use games to explore their identity, ideologies, and connections to culture~\cite{penix2016cultural,cragoe2016rpg}. Thus, the global video game market requires developers to create culturally significant games aligned with the values and needs of players of diverse backgrounds (\textit{Game Culturalization})~\cite{o2013game,pyae2018understanding}. 

Games often incorporate culture through language, music, landscapes, narrative, and aesthetics. ``Pokémon GO,'' players can choose their preferred language and localize non-player characters' (NPCs) names ~\cite{pyae2018understanding}, representing culture through \textit{language's} natural incorporation of symbols and ideologies directly tied to the society from which it developed~\cite{kramsch2014language,wardhaugh2021introduction}. ``Just Dance'' incorporated various \textit{music} from various cultures, including American, Korean, and Indian~\cite{pyae2018understanding}. Games' virtual environments  often use \textit{cultural landscapes and heritage}, that reflect history, values, and connection to nature~\cite{UNESCO2024landscape}, by taking inspiration from important cultural landmarks ~\cite{pyae2018understanding}. The  aesthetics of characters can reflect cultural \textit{fashion}~\cite{pyae2018understanding}, a symbolic and functional form of cultural expression~\cite{ryan2010cultural}. Games also reference culture through their narratives~\cite{cragoe2016rpg,loban2023embedding}; ``World of Warcraft's'' narrative borrows from Sir Thomas Mallory's \textit{Le Morte D'Arthur}~\cite{kristen2014wow}, a work of medieval literature based on British mythology.

Although culture is both tangible and intangible, it shapes human behaviour~\cite{hall1973silent}. Developers often unintentionally embed their own cultural values into their designs~\cite{pyae2018understanding}. ~\citet{bogost2010persuasive} noted that games can effectively employ culture in their mechanics to influence player opinions, emphasizing the need to address the tension between psychological and cultural perspectives on player manipulation~\cite[p.~24]{bogost2010persuasive}. Given that cultural representations can impact player engagement~\cite{Pyae2018engagement} and persuade players' beliefs~\cite{penix2016cultural,anable2018playing}, it is important to recognize that these representations, while valuable for enhancing player experience, have the potential to inadvertently facilitate deceptive game design. This problem is crucial, as deceptive practices amplified by cultural elements can be harder for players to resist, and risk leading to problematic instances of cultural appropriation by developers seeking to replicate this profitable formula. 
%\st{we hypothesize that these representations can also be exploited to enhance deceptive game design, making them harder for players to resist.} 
Due to the lack of research on this topic, we encourage the exploration of cultural representations as a direction for the future deceptive game design research.

\begin{table}[!t]
\centering
\caption{Definitions of deceptive game design patterns discussed in this paper, adopted from~\citet{zagal2013dark}.}
\label{tab:game-deceptive-design}
\resizebox{\columnwidth}{!}{%
\begin{tabular}{@{}llp{0.65\columnwidth}@{}}
\toprule
\textbf{Category} & \textbf{Pattern} & \textbf{Definition~\cite{zagal2013dark}} \\ \midrule
Temporal& Grinding & Patterns that require players to engage in repetitive tasks. \\
Patterns & Daily Rewards & Patterns that incentivize players to revisit daily by offering escalating rewards. \\ 
& Playing by Appointment &  Patterns that forces players to play based on the game's schedule. \\
&Infinite Treadmill & Patterns that constantly expand the game to prevent players from completing. \\
\\
Monetary & Premium Currency & Patterns that conceal the true cost of items using additional exchange rates.\\
Patterns & Pay to Win & Patterns that provide advantages to players who spend money.\\
& Gambling / Loot Boxes & Patterns that incentivize players to spend money by randomizing the rewards.\\ \bottomrule
\end{tabular}%
}
\end{table}

%==============
\section{Example: Interplay between cultural attributes and deceptive game design in Genshin Impact}
\label{sec:discussion}
In this section, we examine a popular gacha game with a focus on its cultural attributes. For clarity, deceptive patterns aligning with~\citet{zagal2013dark}'s definitions (see~\autoref{tab:game-deceptive-design}) are \textit{italicized} (e.g.,~\codestyle{Premium Currency}), and game design elements are in monospace (e.g.,~\mechanics{Lantern Rite}) and demonstrated in a glossary in~\autoref{tab:glossary}.

%==========================
%\hilda{talk about game info and deceptive game designs present in the game}
``Genshin Impact'' is an open-world action role-playing game (RPG) where players control characters with elemental abilities, explore a fantasy world, experience storylines, and complete quests. In 2022, it was the highest-grossing mobile game with \$1.56 billion USD revenue primarily from East Asian countries such as Japan (19.2\%), Singapore (13.5\%), Hong Kong (12.4\%), and South Korea (10.2\%)~\cite{statista2022highest,statista2024genshin}. % compared to other areas of the world such as the United States (6.0\%), Canada (4.2\%), and France (1.6\%). 
The game uses a gacha system, where players can purchase \codestyle{Premium Currency} for randomized ``wishes'' to acquire new characters and weapons, and to repeatedly acquire duplicates to fully unlock their abilities (\codestyle{Gambling/Loot Boxes}). These characters and weapons enhance the gameplay experience and team compositions (\codestyle{Pay to Win}). In addition, the game lures players by rewarding them for completing daily, weekly, and monthly quests that involve repetitive tasks (\codestyle{Daily Rewards}, \codestyle{Infinite Treadmill}, \codestyle{Grinding}).

%==========================
%\hilda{talke about cultural attributes in general}
Genshin Impact's greater appeals arise from its incorporation of cultural attributes that are familiar and engaging to East Asian players. For example, the deeper philosophical aspects of the characters' cultural identities that shape players' perceptions are often conveyed in Chinese but missed in the English translation~\cite{zhou2024integration}. Additionally, the \mechanics{Liyue} region in the game closely resembles landmarks such as Huanglong Scenic Area and Guilin in China. Players collect and cook recipes from various cultures, including \mechanics{Rice Buns} (China), \mechanics{Fish and Chips} (United Kingdom), \mechanics{Taiyaki} (Japan), and \mechanics{P\^{a}te de Fruit} (France). The game also holds festivals such as \mechanics{Moonchase Festival} and \mechanics{Lantern Rite} that reflect China's cultural traditions. Music in the game, such as ``The Divine Damsel of Devastation,'' has increased global interest in Chinese opera~\cite{cai2023study}. Japanese culture is also incorporated in the game's character design and backstories. An example is \mechanics{Raiden Shogun}, whose appearance and narrative blend traditional Japanese aesthetics with symbolic elements and historical references~\cite{agung2023preliminary}. %Her attire, a kimono adorned with a Mitsudomoe emblem, and her sakura flower hair ornament draw direct connections to Shinto shrines and the Japanese God of War~\cite{agung2023preliminary}. The nation she rules, \mechanics{Inazuma}, and her storyline were based on 16th-century Japan~\cite{agung2023preliminary}.  
These aesthetic, narrative, and mechanic choices can be seen as representations of Chinese and Japanese ``cultural heritage'', a concept that describes both the tangible objects and ``intangible attributes'' that represent ``shared practices'' and identity shared by a community and ``can be transferable between generations''~\cite[p.~3--20]{loban2023embedding}. The incorporation of these cultural representations enhances player engagement and creates a deeper emotional connection to the gameplay environment, as evidenced by newsletters hailing the game as a ``rejuvenation'' of Chinese culture~\cite{medium2024genshin} as well as its recognition as the winner of the Best Mobile Game award at The Game Awards 2021~\cite{2021TGA}. 

\begin{table*}[!ht]
\centering
\caption{Genshin Impact Items Glossary. We only included items discussed in~\autoref{sec:discussion}.}
\label{tab:glossary}
\resizebox{\textwidth}{!}{%
\begin{tabular}{cccccc} 
\toprule
Gaming & Raiden Shogun & Orchid's Evening Gown & Rice buns & Fish and Chips & Taiyaki \\ 
\midrule
\includegraphics[width=0.15\textwidth]{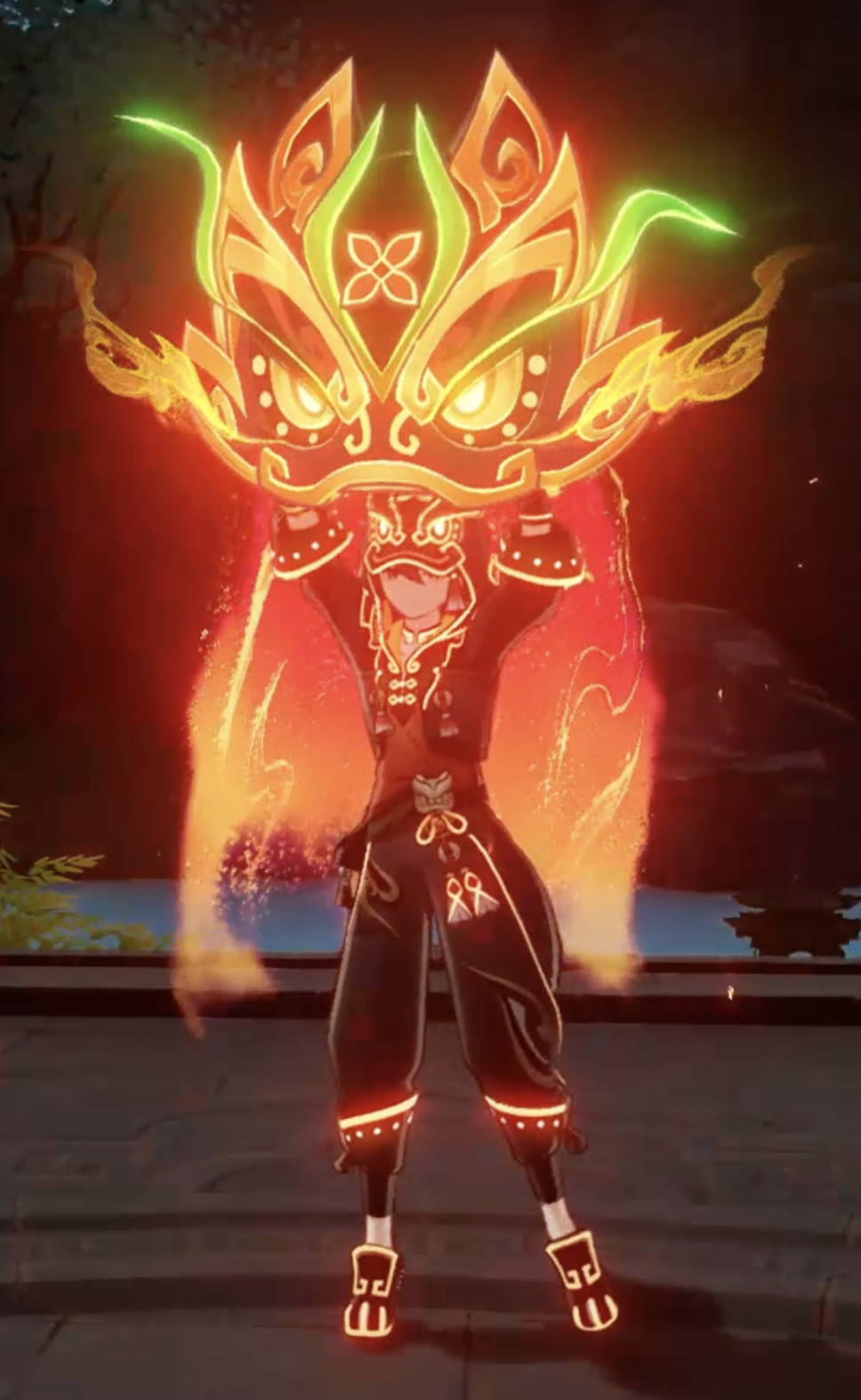}        
        & \includegraphics[width=0.15\textwidth]{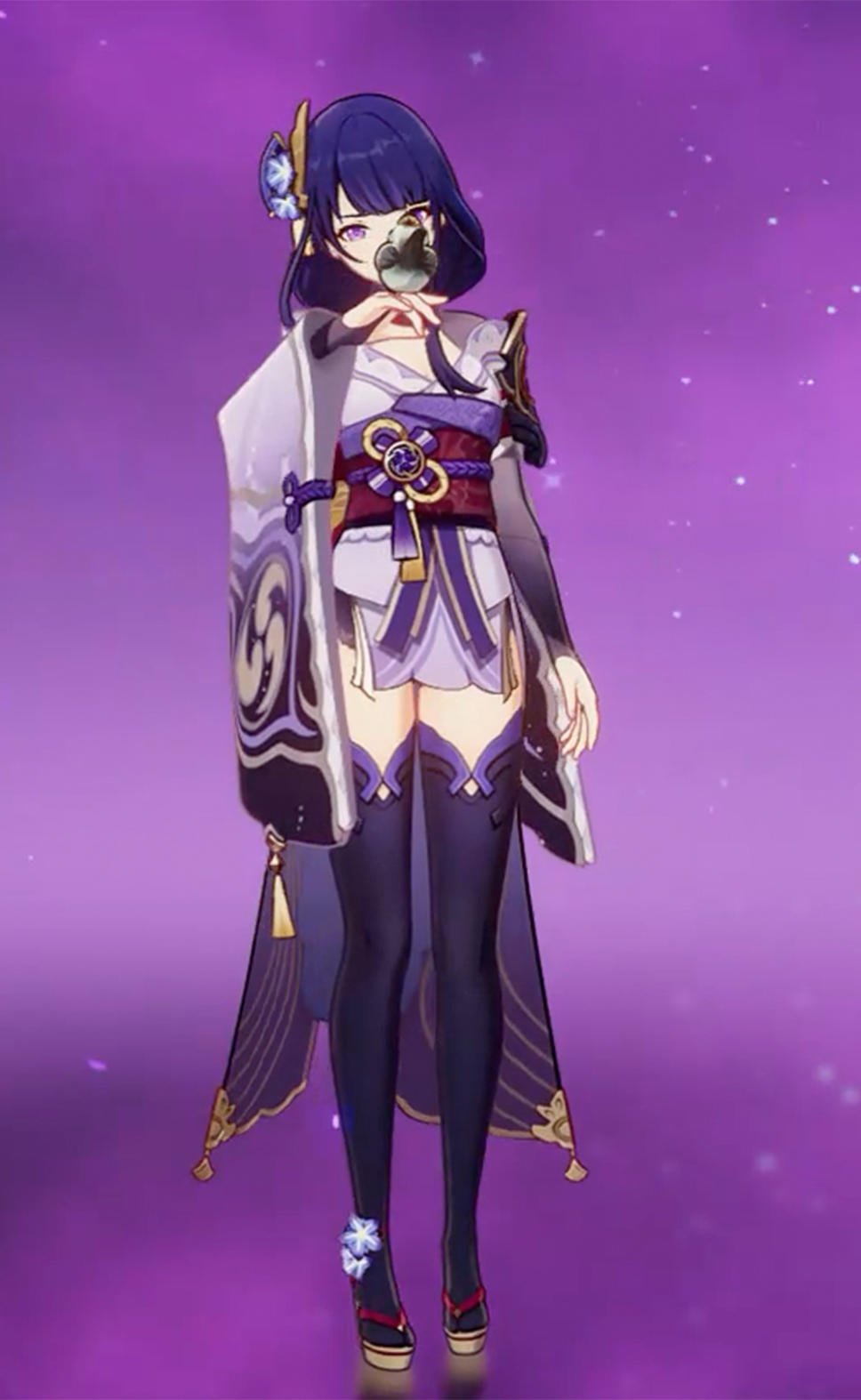}              
        & \includegraphics[width=0.15\textwidth]{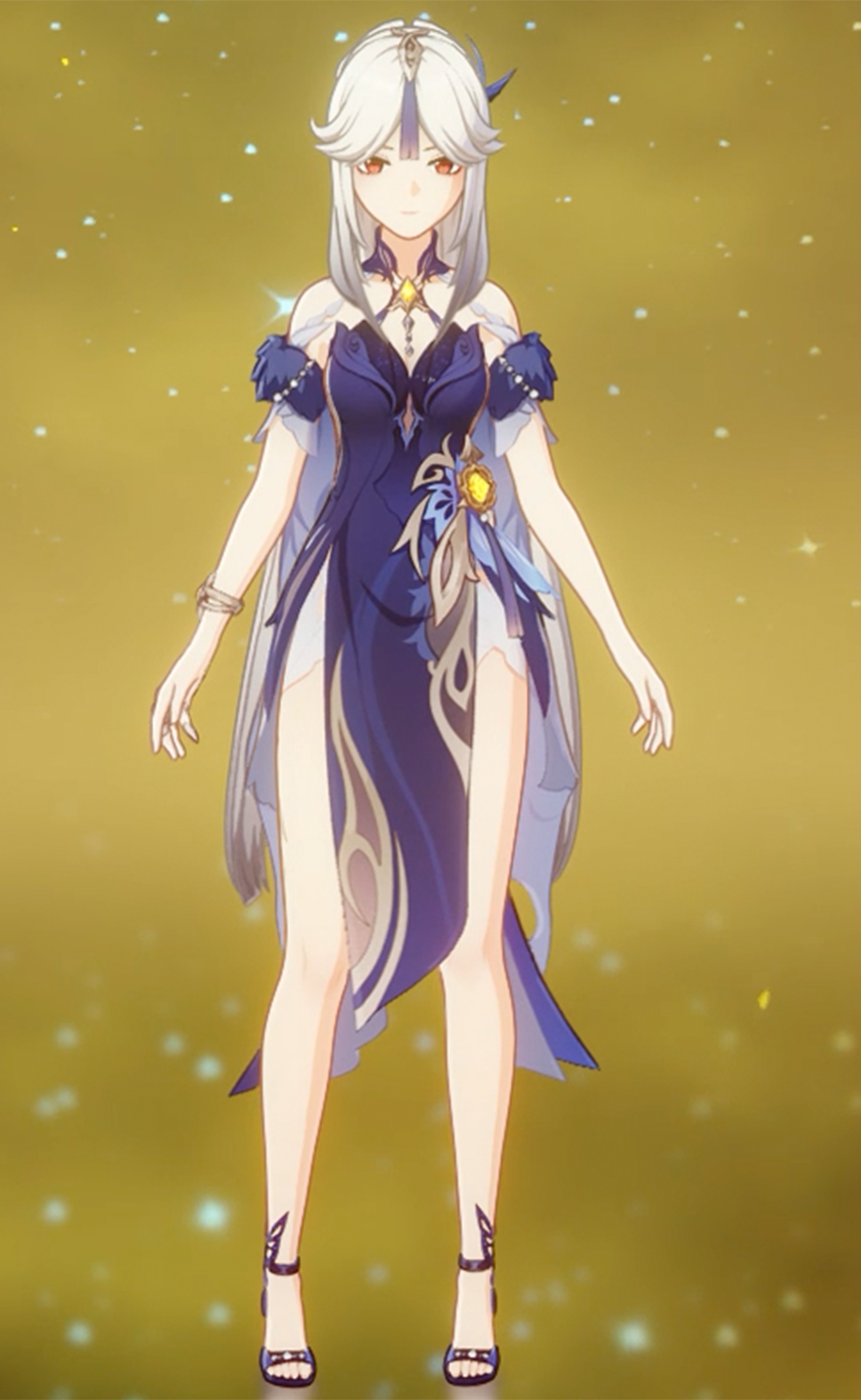} 
        & \includegraphics[width=0.15\textwidth]{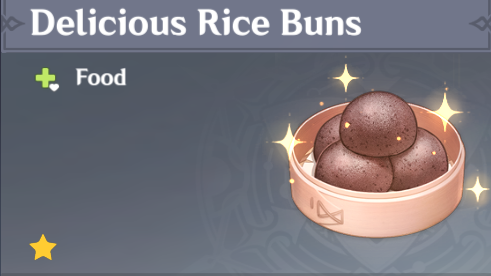}
        & \includegraphics[width=0.15\textwidth]{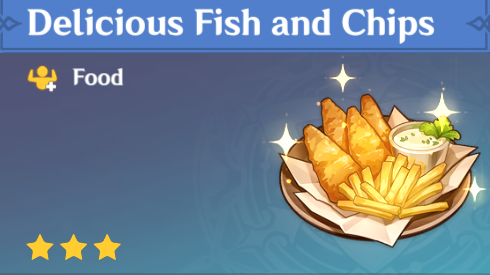}
        & \includegraphics[width=0.15\textwidth]{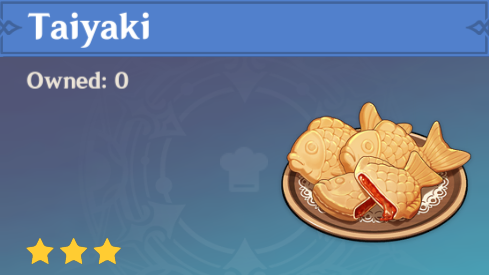} 
%        & \includegraphics[width=0.15\textwidth]{PatedeFruit.png} 
\\ 
\bottomrule
\end{tabular}
}
\end{table*}

%wears a sakura flower and a violet-coloured fan as hair accessories, as well as a kimono with a red and gold ribbon and a Mitsudomoe emblem. This design references a traditional Japanese-style Yukata and is associated with Shinto shrines and the Japanese God of War~\cite{agung2023preliminary}. 

%==========================
%\hilda{talk about how cultural attributes may amplify deceptive game design. The first example is "impulsive purchase" and the second example is "grinding"}

%first example starts here --- impulsive purcase
Nevertheless, many cultural attributes in Genshin Impact serve as a `smokescreen' for the game's deceptive practices to incentivize impulsive purchases. For instance, the character \mechanics{Gaming} released during China's Spring Festival, wears a lion dance costume and has fire-elemental abilities that reflect the festival traditional lion dance performance. This alignment of \mechanics{Gaming}'s design with the Festival traditions of acquiring good fortune has the potential to increase players' susceptibility to the gacha system's allure, leading them to obtain \mechanics{Gaming} for both enjoyment and cultural significance, and even repetitive spending to unlock all his abilities. Similarly, characters closely relevant to Japanese culture, such as \mechanics{Raiden Shogun}, have the potential to promote excessive spending from Japanese players or those obsessed with this culture. 
%second example starts here -- the potential to lead to excessive grinding and play the game based on event schedule
Additionally, many cultural attributes amplified the success of temporal deceptive design practices. For example, in 2022, the game released \mechanics{Orchid's Evening Gown}, a character costume obtainable for free during the limited-time event \mechanics{Fleeting Colors in Flight}. This costume's design resembles a qipao, a high-necked, form-fitting dress with intricate floral embroidery and elegant flowing sleeves, which reflects the sophistication of traditional Chinese fashion. Its representation of traditional cultural significance compels players to grind repetitive limited-time tasks (\codestyle{Grinding}, \codestyle{Playing by Appointment}).

%==============
\section{Future Research Directions}
\label{sec:future-research}

The Genshin Impact example illustrates that, while incorporating cultural attributes such as language, music, arts, and cultural landscapes can enrich the gameplay experience and promote cultural appreciation, their misuse can amplify the success of deceptive practices and be ethically questionable, if not overtly problematic, given the game's ease of access and popularity among younger players. Considering the game's rating of ages 12 and up and its lack of a robust age verification or parental consent mechanisms~\cite{parentzone2024what}, the unfettered access to in-game purchases exposes minors to its addictive gacha mechanics~\cite{zagal2013dark}. Additionally, the intersection of cultural attributes and deceptive game design practices blurs the line between celebration and exploitation, raising questions about the responsibility of game developers in preserving the integrity of cultural representations. Moreover, the problematic use of cultural attributes may disproportionately affect players from cultures being represented in games as they might be more susceptible to the manipulation, particularly if they have strong cultural connections.

The issue of cultural attributes potentially amplifying deceptive game design practices is not limited to Asian-inspired games. For instance, ``Assassin's Creed: Valhalla,'' set in Viking-era England, uses Norse mythology, historical references, and a `neomedieval' setting to enhance its loot box system and time-limited events, potentially exploiting players' interest in and connection to Viking culture to drive engagement and purchases. This is particularly problematic not just because of the wide-spread interest, sense of nostalgia, and connection to Viking myth and Nordic culture over the last decade~\cite{Andersen2019,strand2021viking}, but also given the steady rise in the misappropriation of Viking `mythos' and history by white supremacists to justify their hatred, violence, and xenophobia~\cite{sodergren2022viking}. Therefore, we argue that there is an urgent need for game developers, ethicists, researchers, and policymakers to investigate these issues thoroughly and establish guidelines to balance cultural representation with player protection, ensure respectful use of cultural attributes in games, and prevent their exploitation as deceptive design facilitators.

Our discussion, while offering a preliminary overview, lacked rigorous research procedures and thorough investigation. The assessment of whether Genshin Impact's use of cultural attributes in deceptive game designs truly resulted in player harm and negative experience requires an in-depth study of the general player population's interactions, perceptions, and experiences~\cite{hhadan2024ow2}. Further, as our research team is diverse, multi-ethnic and multi-racial, including individuals from East Asia, our discussion on Genshin Impact is rooted in our team's direct connection to East Asian culture, experience playing this game, and expertise in deceptive design and cultural representation and impact in games. This perspective, while providing valuable insight into the impact of Asian cultural attributes in Genshin Impact, also highlights the need for diverse cultural viewpoints to examine how games use various cultural attributes to potentially amplify deceptive practices. To comprehensively explore the role of cultural attributes in deceptive game design, we propose three possible research directions: \textit{First, what cultural attributes used in game design might consequently facilitate deceptive game design?} This question calls for a thorough investigation to identify and classify which cultural attributes are used for deceptive purposes. Research in this direction can involve analyzing existing games and a broader spectrum of cultural representations in games, interviewing game designers, and conducting player surveys to understand how cultural attributes are currently being used in both positive and manipulative ways. \textit{Second, in what ways do these cultural attributes influence the impact of deceptive game design on players?} Building on the first research direction, studies can also conduct comparative playtesting on different game versions with and without deceptive cultural attributes to measure player in-game purchasing, time-spending, and emotional and psychological responses. \textit{Third, with the evolution of immersive technology (VR) and AI, what are the cultural implications for deceptive game design?} Research has shown that immersive technology features, such as spatial display and interaction with digital objects, multi-modality and realistic virtual experience and sensations, and the vast amount of data from body-tracking sensors, can amplify user manipulation~\cite{hadan2024deceived,krauss2024what}. We suggest that research explores how these technologies can be used to create more immersive and personalized experiences with cultural attributes, while mitigating the potential for amplified manipulation. We believe that investigating these research directions will foster a deeper understanding of the issues of deceptive game design from diverse cultural perspectives and create games that are ethical, entertaining and engaging for players around the globe.

%==============
%\section{Conclusion}
%\label{sec:conclusion}
%\input{04-Conclusion}

%\begin{acks}
%Thank you to the chairs and reviewers for their insightful feedback, which helped us to improve the quality of this manuscript. We also thank graduate researcher Derrick Wang for his support in resolving technical issues during our paper formatting. Screenshots in this manuscript were from the Genshin Impact game and fall under fair use. Any opinions, findings, and conclusions or recommendations expressed in this material are those of the author(s) and do not necessarily reflect the views of the University of Waterloo. 
%\end{acks}

\bibliographystyle{ACM-Reference-Format}
\balance
\bibliography{02-Preferences}

\end{document}